# Relativistic spherical symmetries

A. LOINGER

Dipartimento di Fisica, Università di Milano

Via Celoria, 16 − 20133 Milano, Italy

**Summary.** − The observational "black holes" are quite different objects from the theoretical black holes.

**1.** − The spacetime interval

$$(1.1) \quad ds^2 = \left[1 - \frac{2m}{f(r)}\right] c^2 dt^2 - f^2(r)[d\vartheta^2 + \sin^2\vartheta \; d\varphi^2] - \left[1 - \frac{2m}{f(r)}\right]^{-1} [df(r)]^2 \;,$$

where $f(r) > 2m$ is any regular function of $r$, $(0 \leq r < +\infty)$, gives the solution to Einstein equations representing the gravitational field generated by a point mass $mc^2/G$ − and the field external to a spherically-symmetrical distribution of matter of the same mass [1], [2]. Let us remark, however, that those astronomical objects whose field is described by equation (1.1) *not* restricted by the (logically imperative) condition $f(r) > 2m$ − as it happens, for instance, if $f(r) \equiv r$ and one ascribes the well-known peculiar meaning to the spatial region $r \leq 2m$ (*black hole*) − are quite fictional things, as it was emphasized by several authors, expressly by Einstein and by Rosen − and recently by the present writer, who proved in particular that a ***continued*** gravitational collapse of a massive body must end in a ***simple*** pointlike mass [3], and ***not*** in a black hole.

Nevertheless, the overwhelming majority of the theoretical astrophysicists are nowadays convinced of the physical existence of the black holes (BH's) [4]. They claim that the observations corroborate their belief. The statement does not correspond to reality. As a matter of fact, the observations detect only very large masses concentrated in very small volumes (in particular, with energetic jets of electrons beaming out from two sides), but no observation has ever detected any *characteristic* property of the BH's, for instance the existence of an "event horizon", or the amazing and amusing relationship between mass and density of a BH [4bis].

This means that the "BH's" discovered by the observational astrophysicists are *quite different things* from the theoretical BH's. (This conclusion regards both standard and Kerr's BH's [3]).

We shall give now a plain interpretation of the observational "BH's".

**2.** − Let us consider the gravity field generated by a homogeneous sphere of an incompressible fluid [5], [5bis].

If we set, according to Schwarzschild [5]:

(2.1) $\quad r^3 = (\kappa \rho_0 / 3)^{-3/2} \{(9/4)\cos\chi_a [\chi - (1/2)\sin 2\chi] - (1/2)\sin^3\chi\}$ ,

$$(0 \leq \chi \leq \chi_a < \pi/2) ,$$

where the suffix *a* denotes a value at the surface of the sphere, $\kappa := 8\pi G/c^2$, and $\rho_0$ is the uniform and constant mass-density of the fluid, in the *interior* region we have

(2.2) $$ds^2 = \left(\frac{3\cos\chi_a - \cos\chi}{2}\right)^2 c^2 dt^2 - \frac{3}{\kappa\rho_0}(d\chi^2 + \sin^2\chi \, d\vartheta^2 + \sin^2\chi \, \sin^2\vartheta \, d\varphi^2) ,$$

and *outside* the sphere (Schwarzschild's $\alpha$ is denoted here with $2m$):

(2.3) $$ds^2 = \left(1 - \frac{2m}{R}\right) c^2 dt^2 - R^2(d\vartheta^2 + \sin^2\vartheta \, d\varphi^2) - \left(1 - \frac{2m}{R}\right)^{-1} dR^2 ,$$

where:

(2.3') $$2m = (\kappa\rho_o/3)^{-1/2} \sin^3\chi_a ,$$

(2.3'') $$R := (r^3 + \rho)^{1/3} ,$$

(2.3''') $$\rho := (\kappa\rho_0/3)^{-3/2} \{(3/2)\sin^3\chi_a - (9/4)\cos\chi_a[\chi_a - (1/2)\sin 2\chi_a]\} .$$

Schwarzschild proved that a sphere of a given attractive mass $mc^2/G$ cannot have a radius $P_a$, "measured from outside" – and coinciding with $R_a = (r_a^3 + \rho)^{1/3}$–, smaller than $(9/8)2m$. (The corresponding value of $\chi_a$ is arccos $(1/3)$). This limit holds also if the mass density is *not* uniform.

**3.** – Clearly, the above results can explain the observational "BH's". In particular, if the radius of the sphere is just equal to $(9/8)2m$, Dolan's results [4bis] are perfectly intelligible, without resorting to the *deus ex machina* BH.

From the physical point of view, the above sphere, which generates a gravity field without singularities, is evidently a more realistic model than the pointlike mass.

**4.** – If $U(\mathbf{x})$ is a given Newtonian potential, the left side $\nabla^2 U$ of Laplace-Poisson equation allows us to evaluate the corresponding mass density $\rho(\mathbf{x})$. Analogously, if $g_{jk}(x)$, ($j,k=0,1,2,3$), is a known metric tensor, the left sides $R_{jk} - (1/2)g_{jk}(g^{mn}R_{mn})$ of Einstein equations allow us to evaluate the corresponding components $T_{jk}$ of the mass tensor. Thus, by varying the metric tensor we obtain *infinite* expressions corresponding to the stresses and to the energy-momentum densities of *infinite* matter distributions. And this by *prescinding* from any phenomenological choice of the mass tensor.

The instance of the spherically-symmetrical distributions of matter is particularly interesting. For a *static* distribution. we can write [1]:

$$(4.1)\ ds^2 = \exp[\nu(r)]\ c^2 dt^2 - r^2 \exp[\mu(r)]\ (d\vartheta^2 + \sin^2\vartheta\ d\varphi^2) - \exp[\lambda(r)]dr^2,$$

and the only components of the Ricci-Einstein tensor different from zero are $R_{00}, R_{11}, R_{22}, R_{33} = R_{22}\sin^2\vartheta$. Remember that there exists a relation among $\lambda$, $\mu$ and $\nu$. By varying these functions we have the field of *any* matter distribution with spherical symmetry. Thus, the interval (1.1) yields the sphero-symmetrical solution of the Einstein equations

$$(4.2) \qquad R_{jk} - \frac{1}{2}g_{jk}(g^{mn}R_{mn}) = 0\ ;$$

in a sense, this is an exceptional case, because there are *infinite* expressions of $R_{jk} - (1/2)g_{jk}(g^{mn}R_{mn})$ that describe the entire fields – internal and external – generated by static, continuous distributions of matter with spherical symmetry. In other words, the left sides of Einstein equations give infinite "geometric" expressions corresponding to the mass tensors of static, continuous, sphero-symmetrical matter distributions. *The simple case of Schwarzschild's sphere* [5], see sect.**2***, is only* **one** *of the possible instances*. And it is physically evident that there are infinite theoretical collapses which end in *compact* matter distributions of *finite* dimensions.

(The research projects concerning the BH's should be examined anew – a painful reflection because it is difficult to give up a magic quintessence or a prodigious phlogiston [6]).

> "In meinem Revier
> Sind Gelehrte gewesen,
> Außer ihrem eignen Brevier
> Konnten sie keines lesen".
> J.W.v. Goethe

**REFERENCES**

[1] Cf., e.g., EDDINGTON A.S., *The Mathematical Theory of Relativity*, Second Edition (Cambridge University Press) 1960, p.94.

[2] See also LOINGER A., "Regular solutions of Schwarzschild problem", Nota Breve subm. to *Nuovo Cimento*.

[3] See, e.g., CELOTTI A., MILLER J.C. AND SCIAMA D.W., *Class. Quantum Grav.,* **16** (1999) A3.

[4] LOINGER A., http://xxx.lanl.gov/abs/astro-ph/0001453 (January 26th, 2000). An interesting study on "gravitational collapse to a small volume" was written in 1964 by McVittie, see MC VITTIE G.C., *Astroph. J.*, **140** (1964) 401. For a sceptical article regarding the physical existence of the BH's (a *rara avis*), see KUNDT W., *Mem. Soc. Astr. It.,* **67**−1/2 (1996) 329.

[4bis] From BECKWITH S., *Newsletter-Space Telescope Science Inst.*, **18** (2001) No.1: "Joseph Dolan at Goddard Space Flight Center uncovered evidence for accretion onto a black hole by examining High Speed Photometer data nearly ten years old and applying a technique to search for the signature of matter disappearing beyond an event horizon (Dolan*,* J. *Bull. AAS*, 197-118.05). The signature in this case was a series of pulses emitted by clumps of gas as they break away from the innermost stable orbit around a black hole − about 3 Schwarzschild radii out − and spiral in along a set of orbits with decreasing radii. […]. Each successive circle creates a pulse of decreasing magnitude, increasing width, and decreasing period until the gas descends below the event horizon and disappears without further trace. The signature is unique, because if the gas encountered a compact surface, such as that of a neutron star, the final pulse would end in a burst of energy as the gas came to a sudden stop. [..…]. Dolan's work found its way to the *New York Times* science section in January of this year [2001], once again stimulating public interest in astronomy".

We have here an interpretation *ad captandum vulgus*, because the disappearance "without further trace" can be simply explained with the reasonable assumption that the landing of the gas happened over the surface of a Schwarzschild compact sphere of a radius equal to $(9/8)2m$, see sects. **2** and **3** *infra*.

[5] SCHWARZSCHILD K., *Berl. Ber.*, (1916) 424; for an English version see http://xxx.lanl.gov/abs/physics/9912033 (December 16th, 1999). See also: WEYL H., *Raum-Zeit-Materie*, Siebente Auflage (Springer-Verlag, *etc*.) 1988, sect.**35**; EDDINGTON [1], sect.**72**.

[5bis] Remark that Schwarzschild's solution for an incompressible sphere satisfies perfectly all the canons of relativity theory: in particular, the objection that in an incompressible fluid the speed of sound would be infinite is quite irrelevant.

[6] See ABRAMS L.S., *Can. J. Physics,* **67** (1989) 919. This paper develops a brilliant criticism of the notion of BH.